\documentclass[superscriptaddress,showpacs,aps,prb,reprint]{revtex4-2}
\usepackage{graphicx}
\usepackage{amsmath}
\usepackage{braket}
\usepackage[utf8]{inputenc}
\usepackage{bm}
\usepackage{hyperref}
\usepackage{color}

\begin{document}
\title{Mixed-spin Heisenberg ladders in a magnetic field}
\author{D. S. Almeida}
\author{A. S. Bibiano}
\affiliation{Laborat\'{o}rio de F\'{i}sica Te\'{o}rica e Computacional, Departamento de F\'{i}sica, Universidade Federal de Pernambuco, 50760-901 Recife-PE, Brazil}
\author{W. M. da Silva}
\affiliation{Laborat\'{o}rio de F\'{i}sica Te\'{o}rica e Computacional, Departamento de F\'{i}sica, Universidade Federal de Pernambuco, 50760-901 Recife-PE, Brazil}
\affiliation{Unidade Acad\^{e}mica de Educa\c{c}\~{a}o a Dist\^{a}ncia e Tecnologia, Licenciatura em F\'{i}sica, Universidade Federal Rural de Pernambuco, 52171-900 Recife-PE, Brazil}
\author{R. R. Montenegro-Filho}
\affiliation{Laborat\'{o}rio de F\'{i}sica Te\'{o}rica e Computacional, Departamento de F\'{i}sica, Universidade Federal de Pernambuco, 50760-901 Recife-PE, Brazil}
\date{\today}
\begin{abstract}

In this work, we study alternating mixed-spin $(s,S)$ Heisenberg ladders in the magnetic field $h$ using density matrix renormalization group and linear spin-wave calculations. The $h$ \textit{versus} interchain coupling $J_\perp$ phase diagram for the $(1/2,1)$ case is investigated in detail. { In particular, we demonstrate the compatibility between the critical line estimates and magnetic ordering by analyzing chains with variable values of $J_\perp$ and of $h$ along the chain, $J_\perp$ and $h$ scans, and considering the usual case of chains with uniform couplings}. The magnetization plateau at 1/3 of saturation magnetization, 1/3 - plateau, is observed for $J_\perp>0$ and in a limited range for $J_\perp<0$. The critical Kosterlitz-Thouless transition point, where the 1/3 - plateau closes, is identified through a finite-size analysis of the transverse spin correlation functions.

\end{abstract}

\maketitle

\section{Introduction}

Quantum magnets exhibit a variety of fascinating physical phenomena \cite{Zapf2014, Giamarchi2008}. In particular, a gap in the energy spectrum of a magnetic system gives rise to a plateau in the magnetization curve, such that a quantum phase transition \cite{sachdev2001quantum} to a gapless phase, with a distinct magnetization, takes place at the critical fields bounding the plateau \cite{Zapf2014,Giamarchi2008}. However, depending on the couplings between the components of the system, it can be driven to a Kosterlitz-Thouless-type transition \cite{Kosterlitz_2016,*nobelkosterlitz} if the gap-closing point is reached along a parameter line that maintains the magnetization fixed. A fundamental feature of the quantum state in a magnetization plateau is that its unit period needs to satisfy the Oshikawa, Yamanaka, and Affleck condition \cite{PhysRevLett.78.1984}. On the other hand, in one dimension, the gapless phases are critical, exhibiting an asymptotic power-law decay of the spin correlation functions. The low-energy physics is captured by the Luttinger model \cite{giamarchi2003quantum}, with the asymptotic behavior of the spin correlation functions characterized by the non-universal Luttinger parameter $K$.  

One-dimensional spin-1/2 ladder models \cite{Ishida1994,Azuma1994,PhysRevB.59.11398,Dagotto_Rice1996} have been fundamental for the investigation of interacting quantum matter in one dimension \cite{Ruegg2008,Hikihara2010,Ward2017,nayak2020}. In particular, a two-leg spin-1/2 ladder has a singlet gapped ground state, with short-range spin correlation functions. On the other hand, depending on the distribution of the spins along the chain and the couplings, the ground state of mixed-spin ladders \cite{PhysRevB.57.398,PhysRevB.62.11725,PhysRevB.63.054432,Ivanov2001,aristov2004b,chen2007a,japaridze2007,chandra2010,qi2016a,ahmadi2022a} can show a ferrimagnetic order, as expected by the Lieb-Mattis theorem \cite{Lieb.Mattis,Tian}. In fact, some interesting features are exhibited by other one-dimensional ferrimagnetic models \cite{Noriki2017,verissimo2023}. In particular, the spin-($1/2,1$) and spin-($1/2,5/2$) alternating spin chains also have a ferrimagnetic ground state and display the $1/3$ - plateau \cite{AlcarazandMa,Brehmer1997,PatiJPCM1997, PhysRevB.55.8894, PhysRevB.57.13610,Maisinger1998} and the $2/3$ - plateau \cite{ReneJPCM2011}, respectively, in their magnetization curves. The role of density-dependent magnon hopping and the magnon-magnon interaction terms in a spin-wave approximation,  and the nature of the edge states were investigated \cite{DaSilva2021} with the help of density matrix renormalization group (DMRG) calculations. Furthermore, in the phase diagram of some anisotropic spin models, the $1/3$ plateau closes in a Kosterlitz-Thouless (KT) type transition \cite{PhysRevB.102.035137}. In fact, the KT transition was also observed in anisotropic ferrimagnetic chains \cite{PhysRevE.100.042127, PhysRevB.99.134408,verissimo2024}. Interacting spin-1/2 trimers with isotropic exchange couplings exhibit a 1/3 magnetization plateau, but do not show a Kosterlitz-Thouless phase transition in their phase diagram \cite{montenegro-filho2022}.

In this study, we used the density matrix renormalization group (DMRG) \cite{Schollwock2005,schollwock2007,Schollwock2011} and linear spin-wave theory from a fully polarized vacuum \cite{WMS_PRB} to investigate an alternating (1/2, 1) ladder in a magnetic field. We examined both positive and negative values of the inter-dimer coupling $J_\perp$. The critical lines that define the fully polarized and 1/3 magnetization plateaus were identified, along with the magnetic correlations within these phases. Additionally, the KT transition point was determined through analysis of the transverse spin correlation functions.    

In Section II, we present the Hamiltonian, discuss its glide symmetry, and explain the methodology used to obtain our results. Section III covers the general aspects of the phase diagram $h$ versus $J_\perp$. In Section IV, we calculate the magnon bands from the fully polarized vacuum. Section V discusses the magnetic ordering observed in the phase diagram and compares the results of $h$ and $J_\perp$ scans \cite{*[] [{, and references therein.}] jiang2023} with those of chains with uniform couplings. The KT transition point is determined in Section VI, and a summary is presented in Section VII.    

\section{Mixed-Spin Heisenberg Ladder, Glide Reflection Symmetry, and Methods}

\begin{figure}
\begin{center}
\includegraphics[width=0.4\textwidth]{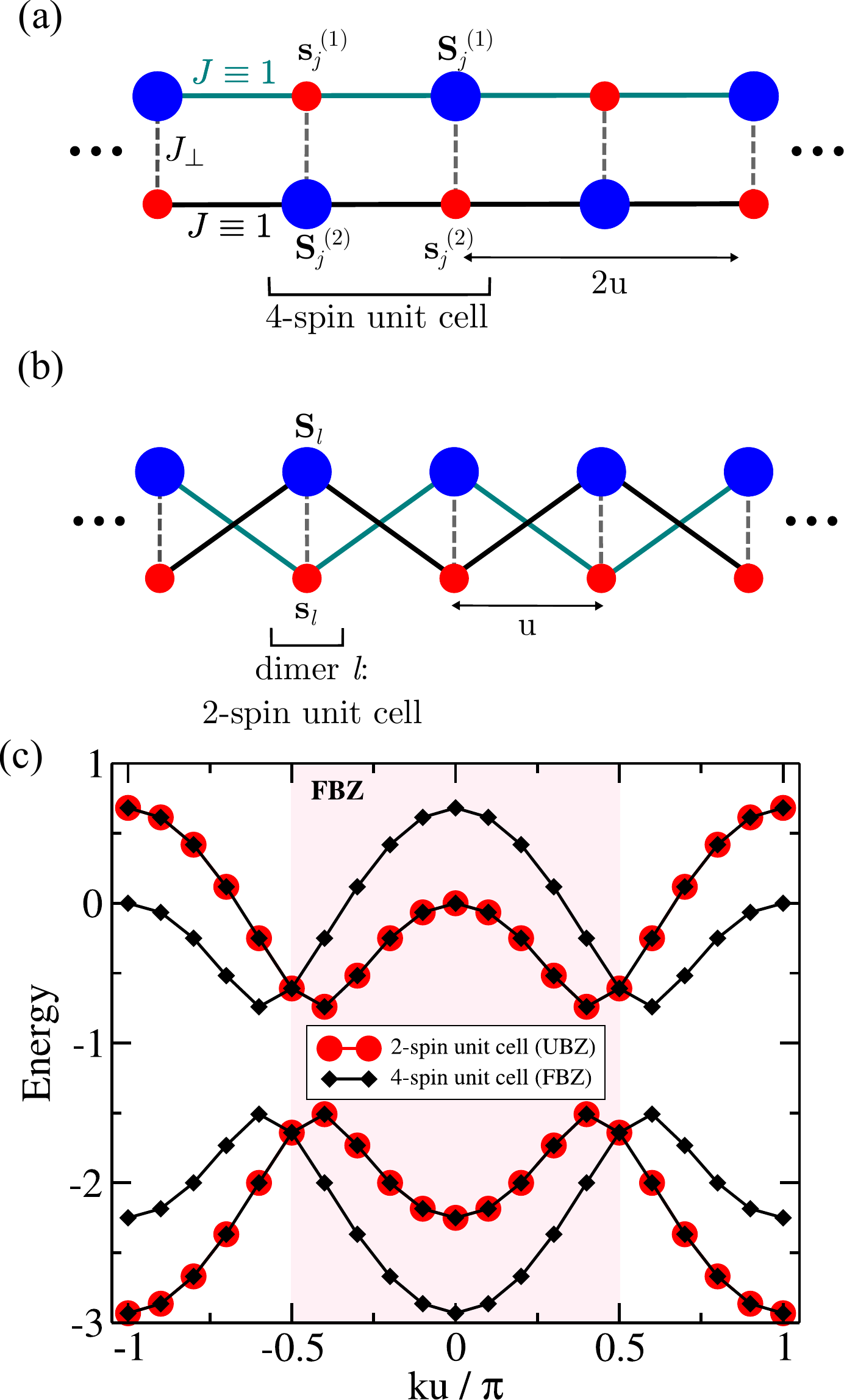}
\caption{
(a) Illustration of the spin-$(s,S)$ ladder Hamiltonian. The unit cell has two spins of magnitude $s$ and two others with magnitude $S$, a 4-spin unit cell with a $2u$ periodicity, where $u$ is the distance between neighboring dimers $l$. The coupling along the legs ($J$) defines the energy unit, and we study the quantum phases of the system as the inter-dimer coupling $J_\perp$ changes. (b) The Hamiltonian with a periodicity of $u$ can become explicit by exchanging the spins at alternate dimers. (c) First Brillouin zone for the model with a 4-spin unit cell, the folded Brillouin zone (FBZ), and 2-spin unit cell, the unfolded Brillouin zone (UBZ), for $J_\perp=-0.5$ in a system with $L=20$ dimers.}
\label{fig:ilust-ham}
\end{center}
\end{figure}

The Hamiltonian of the $(s,S)$-alternating ladder in the presence of a magnetic field $h$ is illustrated in Fig. \ref{fig:ilust-ham}(a) and is given by

\begin{eqnarray}
\mathcal{H} & = & J\sum_{j=1}^{N_c} \Big[ \mathbf{s}^{(1)}_{j} \cdot \mathbf{S}^{(1)}_{j} + \mathbf{S}^{(2)}_{j} \cdot \mathbf{s}^{(2)}_{j} + \mathbf{S}^{(1)}_{j} \cdot \mathbf{s}^{(1)}_{j+1} \nonumber \\
& + & \mathbf{s}^{(2)}_{j} \cdot \mathbf{S}^{(2)}_{j+1} \Big] + J_{\bot}\sum_{j=1}^{N_c} \left[ \mathbf{s}^{(1)}_{j} \cdot \mathbf{S}^{(2)}_{j} + \mathbf{S}^{(1)}_{j} \cdot \mathbf{s}^{(2)}_{j} \right]\nonumber\\
& -& hS^z,
\label{eq:ham4}
\end{eqnarray}
where $\mathbf{s}^{(\alpha)}_j, \mathbf{S}^{(\alpha)}_j$ are the spins of the unit cell $j$ and leg $\alpha$, with $\alpha = 1, 2$ and $j = 1, 2, \cdots N_c$, where $N_c$ is the total number of unit cells. The total spin quantum numbers of $\mathbf{s}^{(\alpha)}_j$ and $\mathbf{S}^{(\alpha)}_j$ are $s$ and $S$ ($S > s$), respectively: $\left[\mathbf{S}^{(\alpha)}_j\right]^2 = S(S+1)$ and $\left[\mathbf{s}^{(\alpha)}_j\right]^2 = s(s+1)$. The superexchange coupling along a leg is $J$ and defines the energy unit, $J\equiv 1$, while $J_\perp$ is the coupling between the two legs. $S^z$ is the z-component of the total spin, and we define $g\mu_B\equiv 1$, where $g$ is the $g$-factor and $\mu_B$ is the Bohr magneton.

We can adopt a spatial representation of the Hamiltonian (\ref{eq:ham4}), illustrated in Fig. \ref{fig:ilust-ham}(a), which is more convenient for any analytical approach exploiting the translation symmetry, such as spin-wave calculations, and that features two spins per unit cell. The Hamiltonian (\ref{eq:ham4}) has a unit cell of size $2u$, where $u$ is the separation between two nearest-neighbor dimers $[s_j^{(1)},S_j^{(2)}]$. The unit cell has four spins, two of size $s$ and two of size $S$, that is, four magnon bands, and a first Brillouin zone of size $\Delta k = 2\pi/(2u)$, where $k$ is the lattice wave vector. However, we also note that the Hamiltonian is invariant under a glide reflection operation \cite{Lee2008,tomic2014,Nica2015}, which is the composite operation of a translation by $u$, followed by the exchange of the ladder leg labels, $1\leftrightarrow2$. This suggests that we can choose a spatial representation such that the Hamiltonian has a reduced unit cell containing the two spins $s$ and $S$, of a dimer, thus two magnon bands, and a first Brillouin zone of size $\Delta k =2\pi/u$. The Brillouin zone of size $2\pi/(2u)$, and a unit cell with four spins, is called the folded Brillouin zone (FBZ); while the Brillouin zone of size $2\pi/u$, and a reduced unit cell with two spins, is the unfolded Brillouin zone (UBZ). In fact, by rearranging the spin indexes of the Hamiltonian (\ref{eq:ham4}) as illustrated in Fig. \ref{fig:ilust-ham}(b), we arrive at a Hamiltonian with a period of $u$ and with two spins per unit cell: 
\begin{equation}
 \mathcal{H} = \sum_{l=1}^{L}(\mathbf{s}_l\cdot \mathbf{S}_{l+1}+\mathbf{S}_l\cdot \mathbf{s}_{l+1})+J_{\perp}\sum_{l=1}^L \mathbf{s}_l\cdot \mathbf{S}_{l}- hS^z,
 \label{eq:ham2}
\end{equation}
where $L=2N_c$ is the number of dimers $(s_l,S_l)$ of the ladder. 

The magnon bands of the Hamiltonian (\ref{eq:ham2}) are calculated in Sec. \ref{sec:swt_fpv} , while the magnon bands of (\ref{eq:ham4}) are obtained in the Appendix \ref{sec:apendice}. As an example, in Fig. \ref{fig:ilust-ham}(c) we show the noninteracting magnon bands of the $(1/2,1)$ chain for $J_\perp=-0.5$ in a system with 20 dimers and periodic boundary conditions: the four bands of the Hamiltonian (\ref{eq:ham4}), the FBZ case, and the two bands of the Hamiltonian (\ref{eq:ham2}), the UBZ case, both considering the fully polarized (FP) state as vacuum.  

The spin-wave calculation is helpful in determining the critical field of the FP plateau and the general properties of other regions of the phase diagram for any value of $s$ and $S$. To obtain precise results for any value of $h$ and $J_\perp$, we use the density matrix renormalization group (DMRG) implementation of the ITensor library \cite{SciPost} for the alternating ladder $(s=1/2, S=1)$ with open boundary conditions. In the DMRG calculations, we consider a maximum discarded weight of $1\times 10^{-10}$ and a maximum bond dimension of 700.

\section{Phase Diagram}

\begin{figure}[!htb]
\begin{center}
\includegraphics[width=0.48\textwidth]{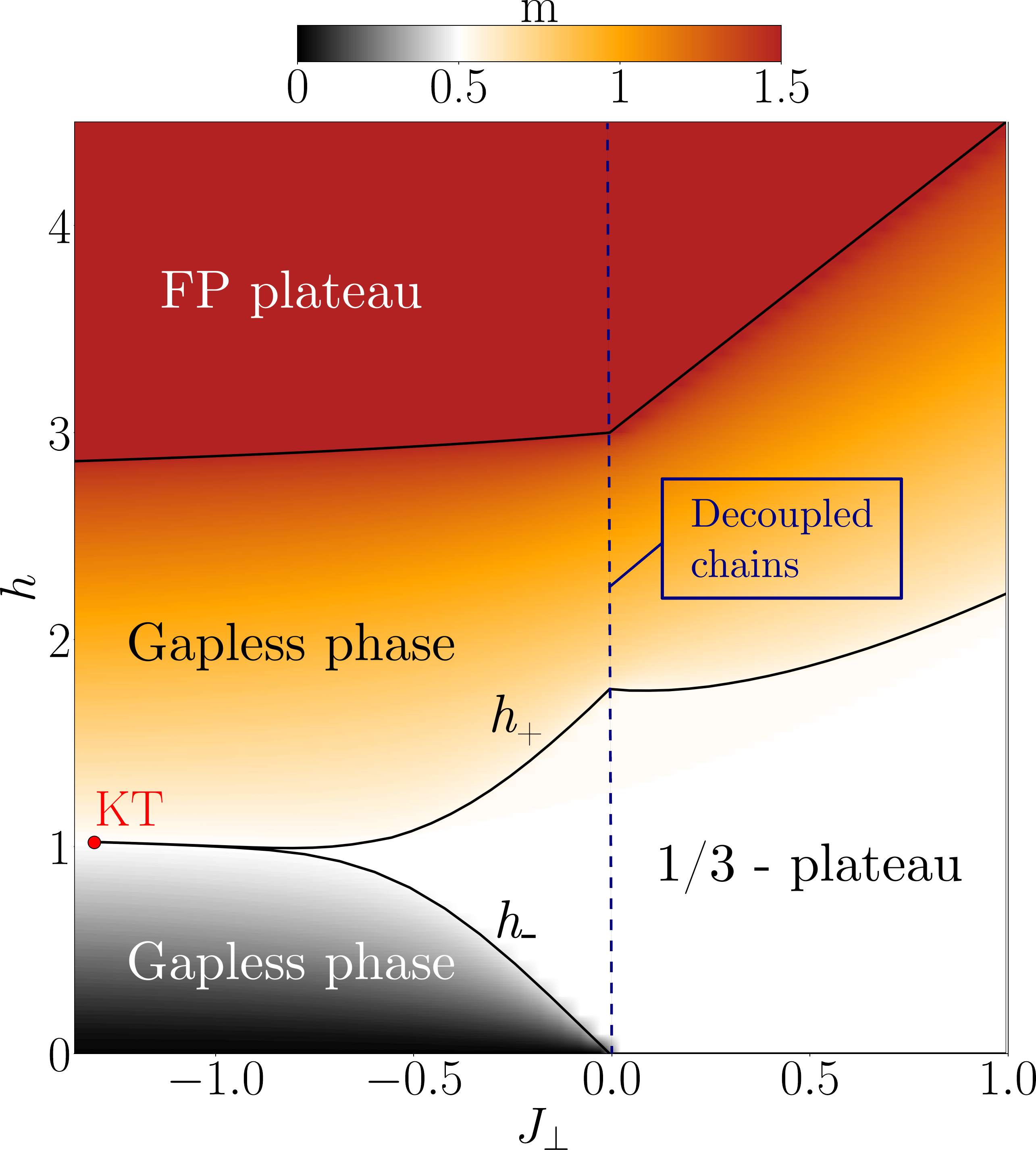}
\caption{DMRG results for the magnetic field $h$ versus rung coupling $J_{\perp}$ phase diagram for the $(1/2,1)$ alternating ladder. The thermodynamic-limit transition lines are estimated from finite-size scale analysis of the magnetization $m$ per dimer as a function of $h$. The color code is the value of $m$ for a system of size $L=100$ dimers. The phase diagram highlights the fully polarized (FP) plateau, the 1/3-plateau, gapless Luttinger liquid phases, and the Kosterlitz-Thouless (KT) transition point. For $J_\perp=0.0$, the $(1/2,1)$ alternating upper and lower chains are decoupled.}
\label{fig:Mixed_Ladder_PhaseDiagram}
\end{center}
\end{figure}

In Fig. \ref{fig:Mixed_Ladder_PhaseDiagram}, we present the phase diagram for the alternating ladder with $s=1/2$ and $S=1$. For $J_\perp=0$, the two alternating $(1/2,1)$ chains are decoupled and have unit cells of size $2u$; see Fig. \ref{fig:ilust-ham}(a). The ground state has a total spin of $1/2$ per unit cell (1/3 of the fully polarized magnetization), as expected by the Lieb-Mattis theorem \cite{Lieb.Mattis}, and the ground state displays a ferrimagnetic long-range order \cite{Tian}. The ground state has a gap $\Delta\approx 1.76$ \cite{Brehmer1997,PatiJPCM1997, PhysRevB.55.8894, PhysRevB.57.13610,Maisinger1998} for excitations increasing the spin, whereas it is gapless for excitations lowering the spin, due to the spontaneously broken rotation symmetry. Thus, in the presence of a magnetic field, the chains exhibit a plateau at magnetization $s^z=(S-s)=1/2$ per unit cell, with extreme critical fields $h_-(J_\perp=0)=0$ and $h_+(J_\perp=0)=\Delta$, where $h_{\pm}$ is given by
\begin{equation}
h_{\pm}=|E[s^z=(S-s)\pm 1,h=0]-E[s^z=(S-s),h=0]|.
\end{equation}
For $s=1/2$ and $S=1$, the plateau occurs at 1/3 of the fully polarized magnetization. The thermodynamic-limit critical lines $h_-(J_\perp)$ and $h_+(J_\perp)$ shown in Fig. \ref{fig:Mixed_Ladder_PhaseDiagram} were obtained through DMRG and finite-size scaling analysis.

For $J_\perp<0$, the exchange couplings do not satisfy the requirements of the Lieb-Mattis theorem. In this case, the ladder exhibits a singlet ground state, $S_{GS}=0$, for $h=0$. However, the 1/3 magnetization plateau persists in the region $J_\perp <0$, but with $h_->0$, and closes at the Kosterlitz-Thouless transition point, at which $h_-=h_+$, in the thermodynamic limit.

The gapless phases are in the Luttinger liquid universality class. The critical line $h_{FP}$, bounding the fully polarized plateau, is calculated precisely through the linear spin-wave theory in Sec. \ref{sec:swt_fpv}, while the Kosterlitz-Thouless transition point is carefully determined from DMRG data in Sec. \ref{sec:kt-transition}. We also discuss in Sec. \ref{sec:kt-transition} the short-range magnetic order in the gapped and gapless phases. 

\section{Spin-wave theory from the fully polarized state}
\label{sec:swt_fpv}

We use linear spin-wave theory from the fully polarized state to determine the critical field $h_{FP}(J_\perp)$. The Hamiltonian (\ref{eq:ham2}) is rewritten in terms of bosonic operators, using the Holstein-Primakoff transformations \cite{WMS_PRB}, such that the $z$-component of the dimer spins $\mathbf{s}_l$ and $\mathbf{S}_l$ are given by

\begin{eqnarray}
 s_{l}^{z} = s - a_l^{\dag} a_l=s - n_{al},  \\  
 S_{l}^{z} = S - b_l^{\dag} b_l=S - n_{bl},
\end{eqnarray}
where $a_l^{\dag}(a_l)$ and $b_l^{\dag}(b_l)$ are bosonic creation (annihilation) operators associated with the spins $s$ and $S$, respectively, of the dimer $l$. Furthermore, the leading-order terms of the ladder operators are given by

\begin{eqnarray}
 s_{l}^{+} & = & (2s)^{1/2}\left(\sqrt{1-\frac{n_{al}}{4s}}\right)a_l\approx (2s)^{1/2}a_l,\nonumber \\
 s_{l}^{-} & = & (2s)^{1/2}a_l^{\dag}\sqrt{1-\frac{n_{al}}{4s}}\approx (2s)^{1/2} a^{\dag}_l, \nonumber \\
 S_{l}^{+} & = & (2S)^{1/2}\left(1-\frac{n_{bl}}{4S}\right)b_l\approx (2S)^{1/2}b_l, \nonumber \\
 S_{l}^{-} & = & (2S)^{1/2}b_l^{\dag}\left(1-\frac{n_{bl}}{4S}\right)\approx (2S)^{1/2}b^{\dag}_l.
\label{eq:bosop}
 \end{eqnarray}

We arrive at the spin-wave Hamiltonian by rewriting the Hamiltonian (\ref{eq:ham2}) in terms of bosonic operators, discarding constant terms, and Fourier transforming:

\begin{equation}
H=\sum_k t_{kk}(a^\dag_kb_k+b^\dag_ka_k)+(\varepsilon_bn_{bk}+\varepsilon_an_{ak}),
\label{eq:sw-ham}
\end{equation}
with $t_{kk}=\sqrt{sS}\left(J_\perp+2\cos{ku}\right)$, and the local potentials
\begin{eqnarray}
\varepsilon_b&=&-s(J_\perp+2)\text{; and}\\
\varepsilon_a&=&-S(J_\perp+2).
\label{eq:localpotentials}
\end{eqnarray}
After diagonalization, we obtain the dispersion relations

\begin{eqnarray}
\omega^{(\pm)}(k)&=&\frac{\varepsilon_a+\varepsilon_b}{2}\pm\frac{1}{2}\sqrt{(\varepsilon_a-\varepsilon_b)^2+4t_{kk}^2}\\
 &=&-\frac{s+S}{2}(J_\perp+2)\nonumber\\
 &\pm&\frac{1}{2}\sqrt{(S-s)^2\left(J_\perp+2\right)^2+4sS\left(J_\perp+2\cos ku\right)^2}.\nonumber\\
\end{eqnarray}

\begin{figure}
\begin{center}
\includegraphics[width=0.45\textwidth]{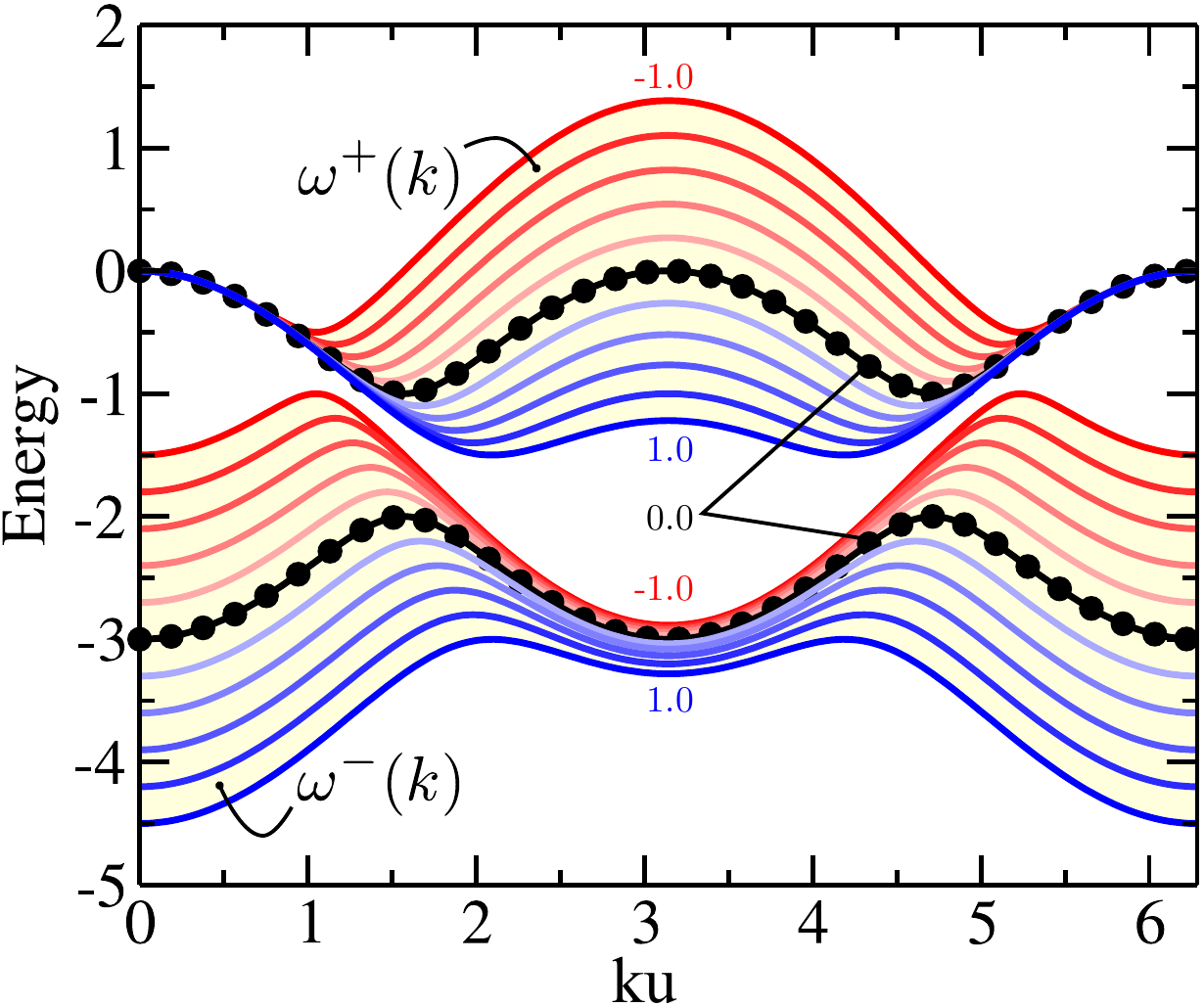}
\caption{Lower [$\omega^-(k)$] and upper [$\omega^+(k)$] free spin-wave magnon bands from the classical ferromagnetic vacuum for the magnetic field $h=0$. In each set, $J_\bot$ ranges from -1.0 to 1.0 with a spacing of 0.2.}
\label{fig:Mixed_Ladder_DispersionRelation}
\end{center}
\end{figure}
In Fig. \ref{fig:Mixed_Ladder_DispersionRelation} we present the two bands for the alternating ladder $(1/2,1)$. 

In the presence of a magnetic field $h$, the bands for that chain are given by
\begin{eqnarray}
\omega_h^{(\pm)}(k)&=&-\frac{3}{4}(J_\perp+2)\nonumber\\
& &\pm\frac{1}{2}\sqrt{\frac{1}{4}\left(J_\perp+2\right)^2+2\left(J_\perp+2\cos{ku}\right)^2}+h.\nonumber\\
\end{eqnarray}
The fully polarized state is stable for $h$ higher than the minimum of the lower band. Since the minimum of $\omega^-$ occurs at $k=0$ for $J_\perp>0$ and at $k=\pi$ for $J_\perp<0$, the exact critical field of the fully polarized plateau is given by
\begin{eqnarray}
h_{FP}(J_\perp)&=&-\omega^{(-)}_{h=h_{FP}}(0)=\frac{3}{2}\left(J_\perp+2\right), ~\text{for $J_\perp>0$};\label{eq:hfp1}\\
h_{FP}(J_\perp)&=&-\omega^{(-)}_{h=h_{FP}}\left(\frac{\pi}{u}\right)=\frac{3}{4}\left(J_\perp+2\right)\nonumber\\
& &+\frac{1}{4}\sqrt{\left(J_\perp+2\right)^2+8\left(J_\perp-2\right)^2}, ~\text{for $J_\perp<0$}.\nonumber\\
\label{eq:hfp2}
\end{eqnarray}

In a first approximation for a many-magnon state, we can consider the magnons as hard-core bosons, with the magnon bands being filled following a spinless fermion restriction. Since the total number of states in the lower band equals the total number of dimers $(1/2,1)$ in the system, the 1/3 - plateau magnetization (one spin flip per dimer from the FP state) is reached when the lower band is full. Thus, the dispersion relations imply the existence of the 1/3 - plateau shown in the phase diagram \ref{fig:Mixed_Ladder_PhaseDiagram}, since the 1/3 - plateau size corresponds to the gap between the lower and upper bands. However,  there is no quantitative agreement with the DMRG data, with the critical fields of the plateau, $h_-$ and $h_+$, far from the exact values. In particular, the gap closes ($h_-=h_+$) at the point $J_\perp=-2$ and $h=0$, which is very different from the numerical value shown in Fig. \ref{fig:Mixed_Ladder_PhaseDiagram}: $J_{\perp,KT}=-1.32$ and $h_{KT}= 1.02$.

\section{Magnetic Order}

\begin{figure}[!htb]
\begin{center}
\includegraphics[width=0.48\textwidth]{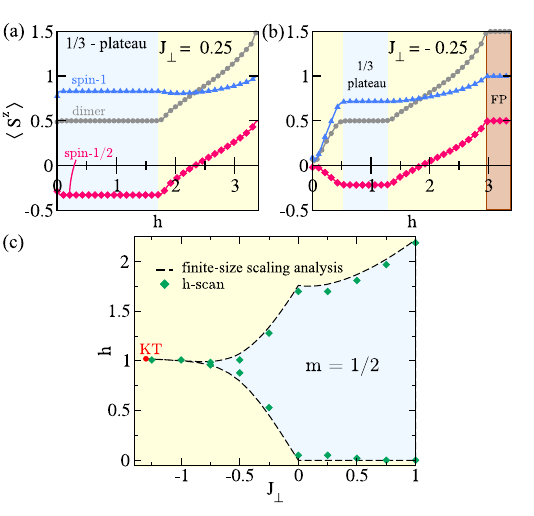}
\caption{Average magnetizations of  spin-1/2  and spin-1 sites, and average dimer magnetization calculated with DMRG from $h$ scans for (a) $J_\bot=0.25$ and (b) $J_\bot=-0.25$. (c) Critical fields estimated from $h$ scans compared with their values obtained from a finite-size scaling analysis of the magnetization per dimer curves.}
\label{fig:magneticorder}
\end{center}
\end{figure}

\subsection{Magnetizations}

In Figs. \ref{fig:magneticorder}(a) and \ref{fig:magneticorder}(b), we show the magnetizations of the spin-1/2 and spin-1 sites and the dimer $(1/2,1)$ for $J_\perp=0.25$ and $J_\perp=-0.25$, respectively. We obtained the data in this figure by performing a $h$-scan calculation \cite{*[] [{, and references therein.}] jiang2023}. Within this approach, we use a ladder chain with a fixed value of $J_\perp$ at all dimers but with a magnetic field that increases linearly from $h=0$ to $h\approx 3.39$ from the left boundary to the right boundary. In particular, we note that $h_{FP}$ agrees with the expressions (\ref{eq:hfp1}) and (\ref{eq:hfp2}): $h_{FP}(J_\perp=-0.25)=2.96$ and $h_{FP}(J_\perp=0.25)=3.38$. In addition, there is a ferrimagnetic orientation between the spin-1/2 and spin-1 sites in the 1/3 - plateau magnetization for both values of $J_\perp$.

Furthermore, the data in Figs. \ref{fig:magneticorder}(a) and \ref{fig:magneticorder}(b) show that the magnon average occupancy of the spin-1/2 sites,
$\braket{n_a}=0.5-\braket{S^z_a}$, is higher than the magnon occupancy of the spin-1 sites, $\braket{n_b}=1-\braket{S^z_b}$, for dimer magnetization between full polarization and $m=1/2$. This tendency is related to the difference between the local potential terms (\ref{eq:localpotentials}): $\Delta \varepsilon=\varepsilon_a-\varepsilon_b=-(J_\perp+2)/2$, which favors the occupancy of the spin-1/2 sites by the magnons. Moreover, since $|\Delta\varepsilon|$ is smaller for $J_\perp = -0.25$ than for $J_\perp = 0.25$, we notice that this imbalance is lower for $J_\perp = -0.25$, compared to $J_\perp = 0.25$. However,  this behavior changes for $J_\perp=-0.25$ and $0<m<0.5$. For that parameter regime, the data show an abrupt magnon occupation of the spin-1 sites, accompanied by a decrease in the magnon occupancy of the spin-1/2 sites. Since in this regime the average magnon occupancy of a dimer,
$\braket{n_{\text{dimer}}}=1.5-m$, is greater than 1, interaction effects become more relevant than the local potential terms.

In Fig. \ref{fig:magneticorder}(c), we compare the estimates from the finite-size scaling analysis with the $h$-scan of the critical fields $h_-$ and $h_+$ for the $1/3$ plateau. We can obtain better results by centering the magnetic field range in the approximate critical fields, while reducing the minimum and maximum values of $h$ in the $h$-scan calculation \cite{*[{}][{, and references therein.}] jiang2023}. We notice the remarkable agreement between the two procedures, although the distinction between $h_-$ and $h_+$ becomes more difficult for the $h$-scan approach as the gap closes in the $J_\perp<0$ region.

\begin{figure}
\begin{center}
\includegraphics[width=0.44\textwidth]{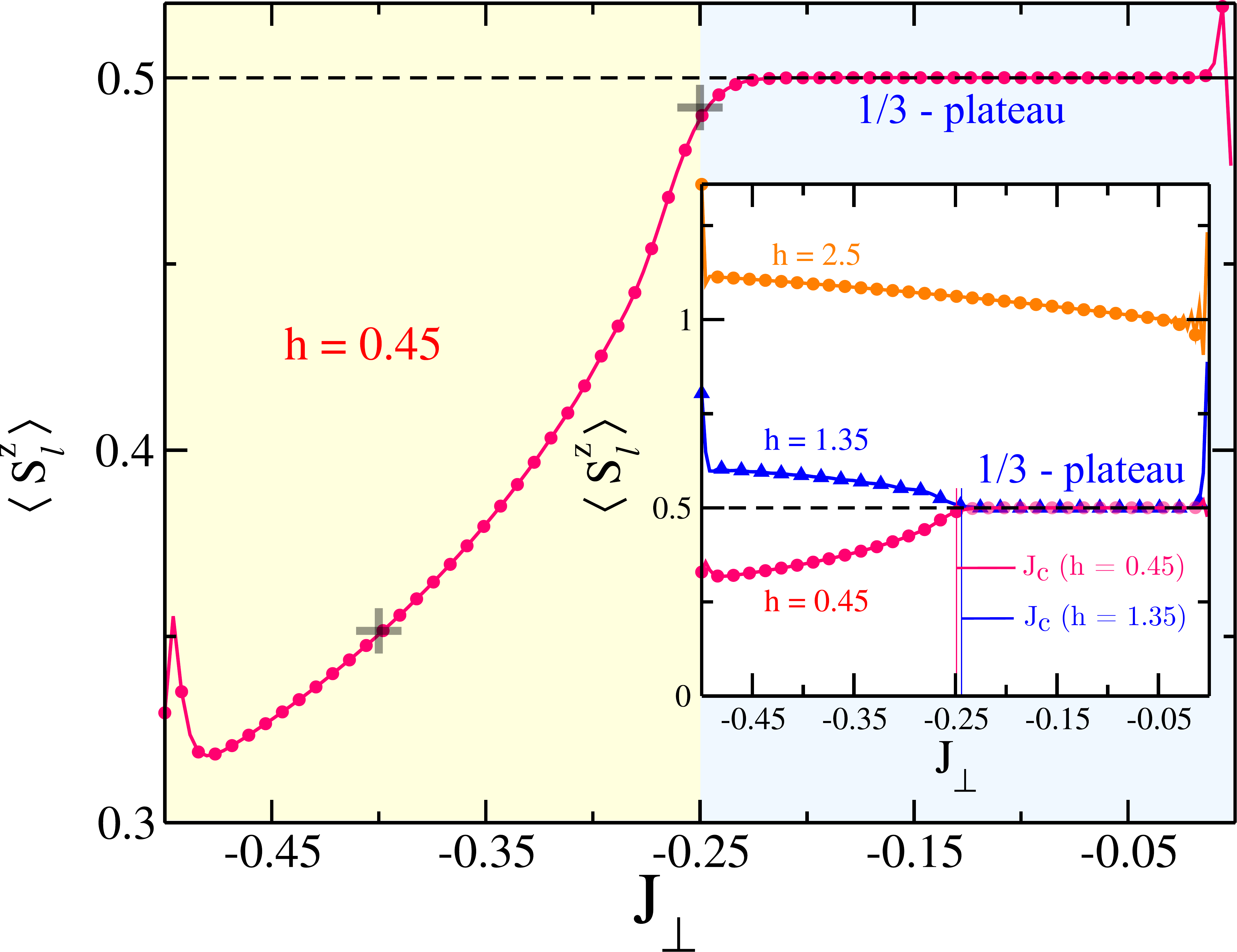}
\caption{Average dimer magnetization calculated with DMRG from a $J_\bot$-scan for $h=0.45$, main figure, and $h=0.45$, 1.35, and 2.5, inset. The critical transition points $J_c$ to the 1/3 - plateau for $h=0.45$ and 1.35 are indicated.}
\label{fig:magneticorder2}
\end{center}
\end{figure}

In Fig. \ref{fig:magneticorder2}, we also present the dimer magnetizations in $J_\perp$-scan calculations
\cite{*[][{, and references therein.}] jiang2023}. In these cases, the magnetic field is the same for all spins, while the value of $J_\perp$ changes linearly from the left to the right boundary. The main figure shows excellent agreement between the magnetization values calculated from a uniform ladder (crosses at $J_\perp=-0.4$ and $-0.25$) and the $J_\perp$- scan for $h=0.45$. Furthermore, the critical point of the 1/3 - plateau shows a tiny departure from the estimated thermodynamic-limit value, while boundary effects are observed. In the inset of Fig. \ref{fig:magneticorder2}, we show the dimer magnetization for three $J_\perp$-scan calculations. For $h=2.5$, the magnetization decreases monotonically and does not reach the 1/3 - plateau ($m=0.5$) in the $J_\perp$ interval exhibited, as expected from the phase diagram \ref{fig:Mixed_Ladder_PhaseDiagram}. On the other hand, the 1/3 - plateau magnetization is observed for $h=1.35$ and 0.45. The comparison between the corresponding critical values of $J_\perp$ in the phase diagram \ref{fig:Mixed_Ladder_PhaseDiagram} confirms a good agreement between the two methodologies. Furthermore, we notice that for $h=0.45$, the magnetization of the 1/3 plateau is reached from lower values of $m$, while for $h=1.35$, the magnetizations are higher than 0.5 for $J_\perp$ below the critical field.

\subsection{Correlations}
\begin{figure}
\begin{center}
\includegraphics[width=0.48\textwidth]{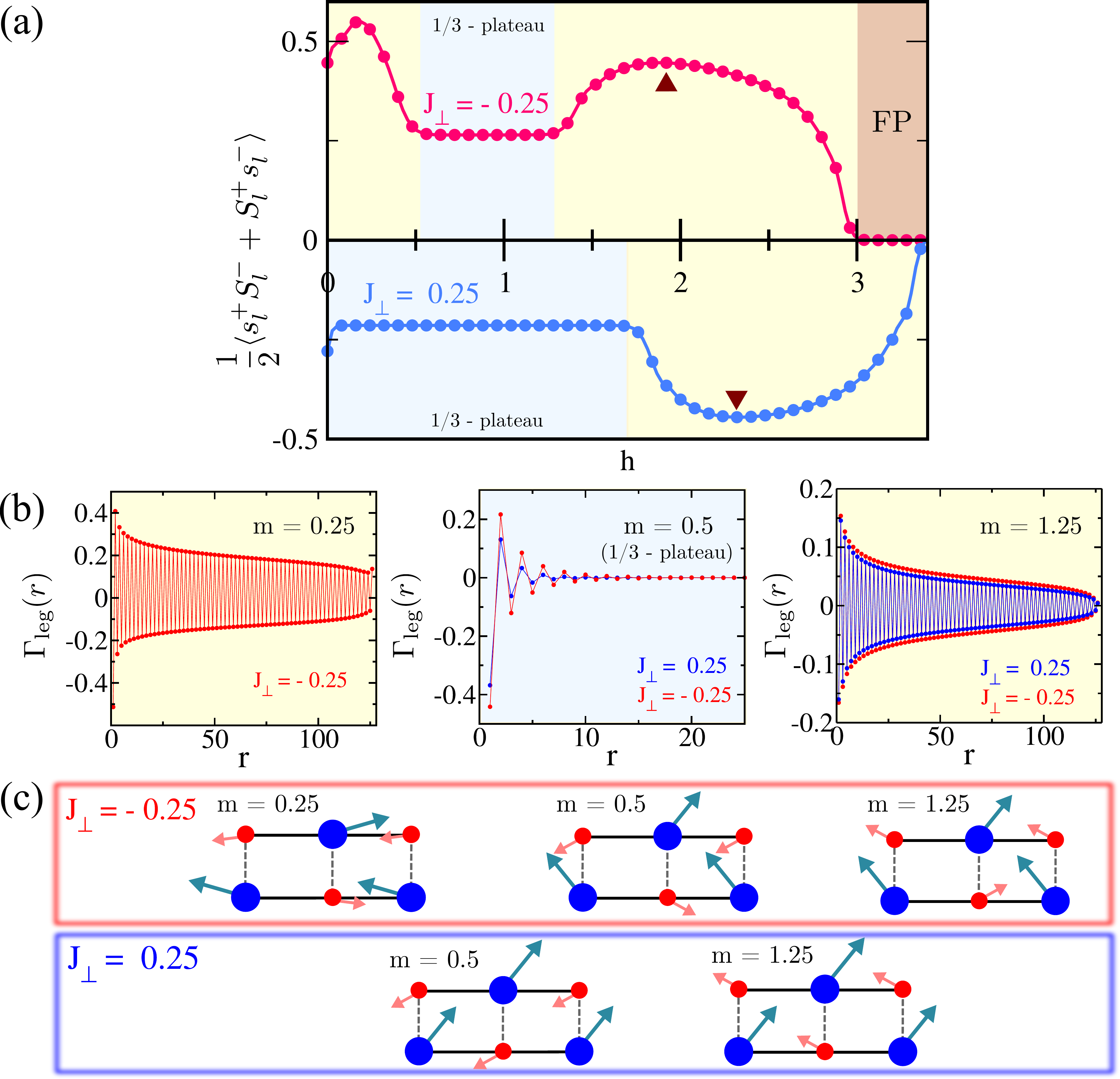}
\caption{(a) DMRG results for the transverse spin correlation function between spins 1/2 and 1  at the same dimer for $J_\perp=-0.25$ and $J_\perp=0.25$ in a $h$-scan calculation for $L=128$. The triangles on both curves mark a local extreme point. (b) DMRG results for the transverse spin correlation function $\Gamma(r)$ for the indicated magnetization per dimer values $m$ and $J_\bot$ for $L=128$. (c) Illustration of the short-range magnetic order for the indicated values of $m$ and $J_\bot$.}
\label{fig:magneticorder3}
\end{center}
\end{figure}

In Fig. \ref{fig:magneticorder3} we show the transverse spin correlation functions between the spin-1/2 and the spin-1 at the same dimer, and the correlation functions along the chain for two typical values of $J_\perp$ in the regions $J_\perp<0$ and $J_\perp>0$. The transverse spin correlation function is defined by
\begin{equation}
 C_{ij}=\frac{1}{2}\langle S^+_{i}S^-_{j}+S^+_{j}S^-_{i}\rangle,
 \label{eq:transvcorr}
\end{equation}
where $i$ and $j$ label the ladder sites. In Fig. \ref{fig:magneticorder3}(a), we observe that for $J_\perp<0$, the transverse correlation function is positive from $h=0$ to the saturation field. In a semiclassical picture, the spin-1/2 and spin-1 in the same dimer have projections in the $xy$ plane that are oriented in the same direction, as sketched in Fig. \ref{fig:magneticorder3}(c). On the other hand, for $J_\perp>0$, the dimer spin projections in the $xy$ plane have opposite orientations, as also sketched in Fig. \ref{fig:magneticorder3}(c). In both cases shown in Fig. \ref{fig:magneticorder3}(a), the correlation does not change in the plateau regions, as expected. Furthermore, we notice that the local extremes of the correlation, marked with a triangle in both
curves, are around the value of $h$ for which the magnetization of the spin-1/2 site, shown in Figs. \ref{fig:magneticorder}(a) and \ref{fig:magneticorder}(b), is null.

The transverse spin correlation function along one of the legs, shown in Fig. \ref{fig:magneticorder3}(b), is defined by
\begin{equation}
 \Gamma_{\text{leg}}(r)_{L}=\braket{C_{ij}}_{|l(i)-l(j)|=r},
 \label{eq:transvcorrleg}
\end{equation}
for a system of size $L$, where $l(i)$ is the dimer index for the site $i$, see Fig. \ref{fig:ilust-ham}(b).
To minimize boundary effects, we consider the spatial average for all pairs of sites in the same leg separated by the distance $r$ in Eq. (\ref{eq:transvcorrleg}). We notice that in the gapless phases ($m=0.25$ and $m=1.25$) the correlation function exhibits the power-law behavior of the Luttinger liquid phase, except for the largest distances due to the open boundaries.
On the other hand, the correlation functions display the exponential decay of a gapped phase for the $m=0.5$ plateau magnetization. For both cases, $J_\perp>0$ and $J_\perp<0$, the transverse spin correlations have an alternating sign along one of the legs, as shown in Fig. \ref{fig:magneticorder3}(c). Finally, the magnetization of the spin-1/2 and spin-1 sites shown in Fig. \ref{fig:magneticorder} can be used to complete the semiclassical picture shown in Fig. \ref{fig:magneticorder3}(c).

\section{Kosterlitz-Thouless transition} 
\label{sec:kt-transition}

\begin{figure}
\begin{center}
\includegraphics[width=0.48\textwidth]{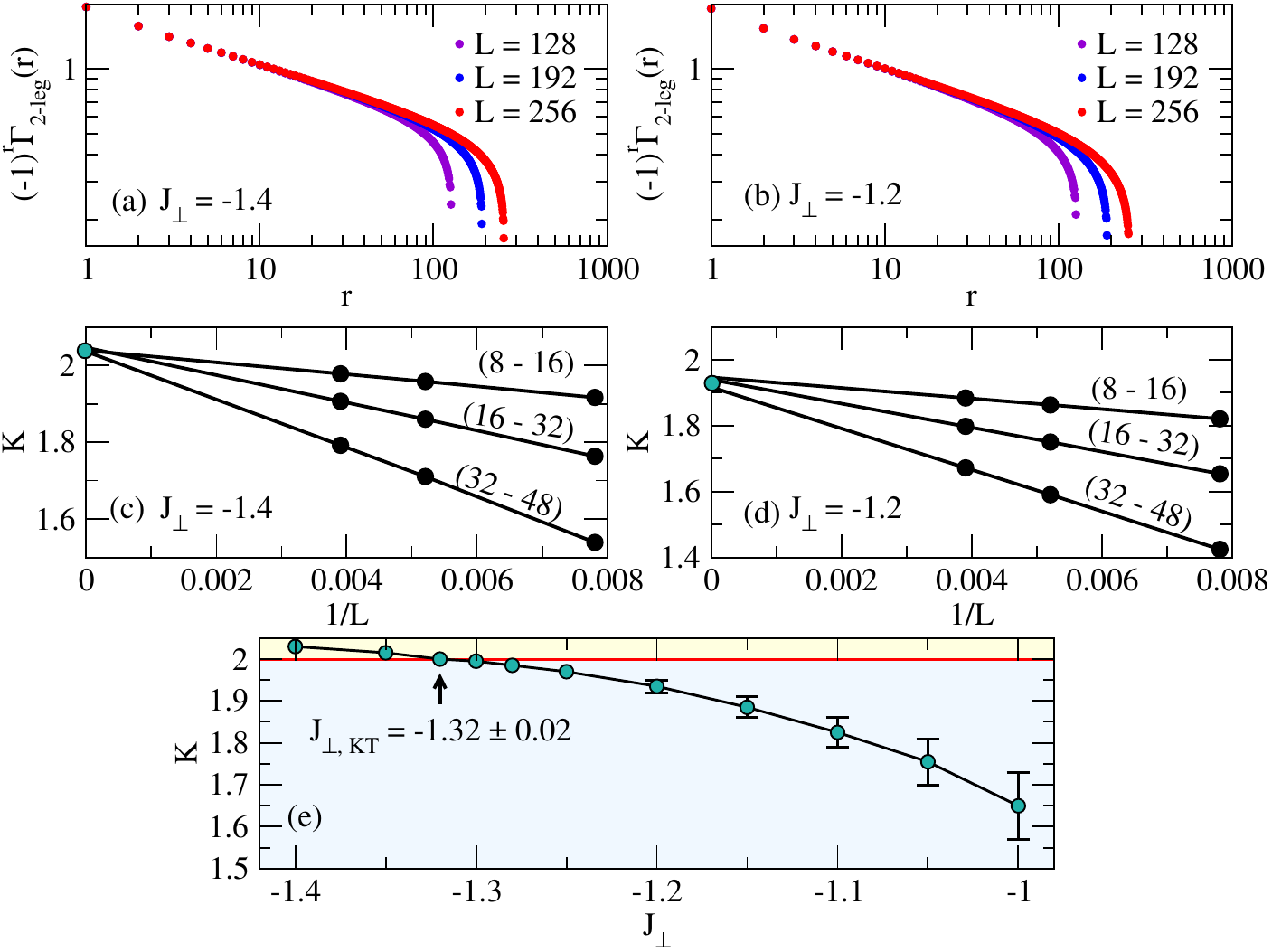}
\caption{{DMRG results for the transverse spin correlation function $(-1)^r\Gamma(r)$, where $r$
is the distance along a leg, at $m = 1/3$ for (a) $J_{\perp}=-1.4$ and (b) $J_{\perp}=-1.2$, and the indicated system sizes $L$.
Luttinger liquid exponent $K$ for (c) $J_{\perp}=-1.4$ and (d) $J_{\perp}=-1.2$ calculated by fitting the transverse correlation function data to the form $1/r^{1/2K}$ through the intervals of distances $8 \leq r \leq 16$, $16 \leq r \leq 32$, and $32 \leq r \leq 48$. (e) Thermodynamic-limit value of $K$ as a function of $J_{\perp}$ near the Kosterlitz-Thouless transition point. We estimate the critical point as $J_{\perp,KT}=-1.32\pm 0.02$.}}
\label{fig:kttransition}
\end{center}
\end{figure}

In the gapless phases, the asymptotic behavior of the transverse spin correlation functions follows a power-law:
\begin{equation}
 \Gamma(r)\sim \frac{1}{r^{1/2K}},
 \label{eq:ll-corr}
\end{equation}
where $K$ is the Luttinger parameter. In the 1/3 plateau, there are 2 bosons per unit cell, thus an integer filling. In these cases, $K=2$ at the Kosterlitz-Thouless transition point \cite{Cazalilla_2011}. To determine the Kosterlitz-Thouless transition point, $J_{\perp, \text{KT}}$, we can fix the magnetization at its value for the 1/3 -- plateau, $m=1/2$, and change the value of $J_\perp$ to localize the point at which $K=2$.

The procedure is more complex for finite-size systems \cite{Kuhner,Montenegro-Filho2020,almeida2023a}, due to boundary effects and the exponentially vanishing gap near the KT point. We calculate the correlation through Eq. (\ref{eq:ll-corr}) in systems of size $L=128,~192,~\text{and }256$ for a given range of $J_\perp$ in the approximate vicinity of the transition point, as exemplified in {Figs. \ref{fig:kttransition}(a) and \ref{fig:kttransition}(b) for $J_\perp=-1.4$ and $J_\perp=-1.2$, respectively}. To estimate the value of $K$ in the thermodynamic limit, we arbitrarily fix some intervals of distances $r$, in our case [8,16], [16,32], and [32,48], and fit the correlation data for each size $L$ to the expression in Eq. (\ref{eq:ll-corr}). The thus obtained values of $K$ from each interval are extrapolated to the thermodynamic limit, as shown in {Figs. \ref{fig:kttransition}(c) and \ref{fig:kttransition}(d), for $J_\perp=-1.4$ and $J_\perp=-1.2$, respectively}. We estimate the thermodynamic limit value of $K$, and the error, by the interval of values of $K$ obtained as $L\rightarrow\infty$, see {Figs. \ref{fig:kttransition}(c) and \ref{fig:kttransition}(d)}.

We show in Fig. \ref{fig:kttransition}{(e)} $K$ as a function of $J_\perp$ near the KT transition. From this curve, we estimate $J_{\perp,\text{KT}}=-1.32\pm 0.02$, which is the point at which $K$ crosses the line $K=2$. Notice, in particular, that this transition point is consistent with the error bar behavior, since the error increases in the gapped phase due to the finite-size effects and becomes smaller than the symbol size in the gapless phase.

\section{Summary}

This work uses linear spin wave theory and the density matrix renormalization group to investigate alternating isotropic mixed-spin ladder chains, particularly with alternating spin-1/2 and spin-1. These chains exhibit glide symmetry and a two-band $ k $-space representation, considering two spins per unit cell in real space or four bands when considering four spins per unit cell. The phase diagram of the magnetic field $ h $ versus inter-dimer coupling $ J_\perp $ presents two magnetization plateaus: the fully polarized plateau and the 1/3 magnetization plateau. In particular, the 1/3 plateau exists for negative values of $J_\perp$ and closes at $ J_\perp = -1.32 $ in a Kosterlitz-Thouless (KT) type transition. The KT transition point was determined from the transverse correlation function since, at the transition, the Luttinger parameter is $ K = 2 $. The critical fields delimiting the fully polarized plateau are calculated exactly through the magnon dispersion relations, considering the fully polarized state as a vacuum. However, the presence of the 1/3 - plateau is correctly predicted by assuming a hard-core boson approximation and free-spin waves. However, the critical KT point and the plateau sizes obtained from this approach significantly deviate from the exact values. Finally, our results reinforce the effectiveness of the $h$ and $J_\perp$ scans in determining the critical fields of the magnetization plateaus by comparing their results with those of the conventional approach using chains with uniform couplings.

Interesting aspects that deserve further investigation include the effect of disorder \cite{*[] [{, and references therein.}] kanbur2020}, and the coupling between legs in the edge states observed in (1/2,$S$) single chains \cite{DaSilva2021}, particularly in the case of coupled ferrimagnetic alternating ladder systems \cite{ovchinnikov2012a}. 

\section{ACKNOWLEDGMENTS}

We acknowledge the support from Coordenação de Aperfeiçoamento de Pessoal de Nível Superior (CAPES), Conselho Nacional de Desenvolvimento Científico e Tecnológico (CNPq), and Fundação de Amparo à Ciência e Tecnologia do Estado de Pernambuco (FACEPE), Brazilian agencies, including the PRONEX program, which is funded by CNPq and FACEPE, Grant No. APQ-0602-1.05/14.

\appendix
\section{Spin-wave theory for a 4-spin unit cell}
\label{sec:apendice}

The Hamiltonian (\ref{eq:ham4}) can be rewritten up to $\mathcal{O}(S^0)$ in terms of bosonic operators as

\begin{multline}
	\bar{\mathcal H}_0 =  \sum_{j = 1}^{N_c} \Bigg\{\sum_{\alpha=1,2}\left(\varepsilon_a a_j^{\dag(\alpha)} a^{(\alpha)}_{j} + \varepsilon_b b_j^{\dag(\alpha)} b^{(\alpha)}_{j} \right)\nonumber \\
+ \sqrt{sS}\bigg[\sum_{\alpha=1,2}\left(a_j^{\dag(\alpha)} b^{(\alpha)}_j + a_j^{(\alpha)} b^{\dag(\alpha)}_j\right) + b_j^{(1)} a^{\dag(1)}_{j+1}+b_j^{\dag(1)} a^{(1)}_{j+1}\\ + a_j^{(2)} b^{\dag(2)}_{j+1} + a_j^{\dag(2)} b^{(2)}_{j+1}\bigg]\nonumber\\
 + J_{\perp}\sqrt{sS}\bigg[a_j^{(1)} b^{\dag(2)}_{j} + a_j^{\dag(1)} b^{(2)}_{j} + b^{(1)}_{j} a_j^{\dag(2)}	+ b^{\dag(1)}_{j} a_j^{(2)} \bigg] \Bigg\},
\end{multline}
taking $J = 1$,  dropping the constant term $2sS(J_{\perp} + 2)$ and making $h=0$. The labels $j$ are sketched in Fig. \ref{fig:ilust-ham} and $N_c$ is the number of unit cells of size $2u$, which we define as $2u\equiv 1$ in the following.

The bosonic operators can be written in $k$-space as

\begin{align}
a_j^{(\alpha)} &= \frac{1}{\sqrt{N_c}}\sum_k e^{ikj}a_k^{(\alpha)},\\  
b_j^{(\alpha)} &= \frac{1}{\sqrt{N_c}}\sum_k e^{ikj}b_k^{(\alpha)},
\end{align}

such that

\begin{multline*}
	\bar{\mathcal H}_0 =  \sum_{k,\alpha}\bigg[\varepsilon_a a_k^{\dag(\alpha)} a_k^{(\alpha)} + \varepsilon_b b_k^{\dag(\alpha)}b_k^{(\alpha)} + \sqrt{sS}a_k^{(\alpha)} b_k^{\dag(\alpha)} \\ + \sqrt{sS}a_k^{\dag(\alpha)} b_k^{(\alpha)}\bigg] + \sqrt{sS}\sum_k\bigg[  e^{-ik}\left( a_k^{\dag(1)} b_k^{(1)} + a_k^{(2)}b_k^{\dag(2)}\right) \\ +  e^{ik}\left(a_k^{(1)} b_k^{\dag(1)} + a_k^{\dag(2)}b_k^{(2)}\right) \bigg] + \sqrt{sS}J_{\bot}\bigg[a_k^{(1)} b_k^{\dag(2)} + b_k^{(1)} a_k^{\dag(2)} \\ + a_k^{\dag(1)} b_k^{(2)} + b_k^{\dag(1)} a_k^{(2)}\bigg],
\end{multline*}
which in matrix form can be written as

\begin{equation*} 
\bar{\mathcal H}_0 =
\begin{bmatrix}
 a_k^{\dag(1)} & b_k^{\dag(1)} &  b_k^{\dag(2)} & a_k^{\dag(2)} \\
\end{bmatrix}
\tau_k
\begin{bmatrix}
a_k^{(1)} \\
b_k^{(1)} \\
b_k^{(2)} \\
a_k^{(2)} \\
\end{bmatrix},
\end{equation*}

where $\tau_k$ is given by

\begin{equation*}
\tau_k = \begin{bmatrix}
\varepsilon_a & \sqrt{sS}\gamma(-k) & \sqrt{sS}J_{\bot} & 0 \\
\sqrt{sS}\gamma(k) & \varepsilon_b & 0 & \sqrt{sS}J_{\bot} \\
\sqrt{sS}J_{\bot} & 0 & \varepsilon_b & \sqrt{sS}\gamma(-k) \\
0 & \sqrt{sS}J_{\bot} & \sqrt{sS}\gamma(k) & \varepsilon_a \\
\end{bmatrix}.
\end{equation*}

Diagonalizing the matrix $\tau_q$, we obtain the four magnon bands shown in Fig. \ref{fig:ilust-ham}:
\begin{eqnarray}
\omega_k^{a(1)} & = & \frac{(\varepsilon_a + \varepsilon_b)}{2} + \frac{\omega^{+}_k}{2} ,\\
\omega_k^{b(1)} & = & \frac{(\varepsilon_a + \varepsilon_b)}{2} - \frac{\omega^{+}_k}{2} ,\\
\omega_k^{b(2)} & = & \frac{(\varepsilon_a + \varepsilon_b)}{2} + \frac{\omega^{-}_k}{2} ,\\
\omega_k^{a(2)} & = & \frac{(\varepsilon_a + \varepsilon_b)}{2} - \frac{\omega^{-}_k}{2},
\end{eqnarray}
with
\begin{equation}
\omega^{\pm}_k = \sqrt{\left(\varepsilon_a - \varepsilon_b\right)^2 + 4sS\left[J_{\perp} \pm 2\cos(k/2)\right]^2},
\end{equation}
while $\varepsilon_a$ and $\varepsilon_b$ are defined in Eq. (\ref{eq:localpotentials}).

\end{document}